\documentclass[12pt]{iopart}
\usepackage{iopams}
\usepackage{graphics}
\usepackage{graphicx}
\begin{document}

\title[Photon gas with hyperbolic dispersion relations]{Photon gas with hyperbolic dispersion relations}

\author{Morteza Mohseni}

\address{Physics Department, Payame Noor University, Tehran 19395-3697, Iran}
\ead{m-mohseni@pnu.ac.ir}
\begin{abstract}
We investigate the density of states for a photon gas confined in a nonmagnetic 
metamaterial medium in which some components of the permittivity tensor are negative. We study the effect
of the resulting hyperbolic dispersion relations on the black body spectral density. We show that for both of the possible wave-vector space topologies, the spectral density vanishes at a certain
frequency. We obtain the partition function and derive some thermodynamical quantities of the system. To leading order, the results resemble those of a one or two dimensional photon gas with
enhanced density of states.
\end{abstract}

\vspace{2pc}

\section{Introduction}
The density of states is a basic factor governing the physical properties of many statistical systems. This can be seen, for instance, by considering the effect of electronic density of states on
conductivity or other properties of condensed matter systems. Similarly, there are many properties of optical systems which depend on photonic density of states. This fact has been used in
designing photonic devices with desired properties by altering their photonic density of states through a variety of techniques \cite{mor,pel,hug,lod}, and more recently, by using
metamaterials \cite{jac}. The later method relies on the properties of metamaterials, which, roughly speaking, are media with negative indices of refraction in certain frequency ranges.

The propagation of electromagnetic waves in media with negative permittivity and or permeability was studied decades ago by Veselago \cite{ves}. This results, depending on the signs of the
permittivity and the permeability,  in the emergence of so-called evanescent or backward-propagating electromagnetic waves (for a brief review, see e.g., Ref. \cite{smi} and for a more
extensive review see Ref. \cite{rama}). Following the experimental realization of such media \cite{nem,she,ele,grb,grbi}, the subject has attracted a lot of interest in both science and 
technology. One can mention in this regard, on the practical side, the construction of various microwave devices ranging from antennas to band-pass filters and lenses, see e.g.,  Ref. \cite{tan} 
and references therein. On the other hand, from the more formal point of view, metamaterials are considered as arenas in which one can mimic certain theoretical models such as dynamic 
spacetime, string theory D-branes and noncommutativity \cite{xin}, spacetimes with a metric signature transition \cite{smo},  Schwarzschild (anti-) de Sitter black holes \cite{mac}, and motion 
around celestial bodies \cite{cles} .

The present work aims to study the thermodynamics of a photon gas confined within a metamaterial medium, which usually obeys hyperbolic dispersion relations \cite{opt}. This is partly
motivated by the recent interest in relativistic gases with modified dispersion relations. Such modifications have been considered within several contexts, namely,  in Refs. \cite{col,leh,kos,
gho,rom}, in connection with Lorentz invariance violating models within the context of quantum field theory; in Ref. \cite{jacob}, in the context of quantum gravity; in Ref. \cite{liberati} in the
context of high energy astrophysics, in Refs. \cite{gur,cos} in the framework of noncommutative field theories, in Ref. \cite{mini}, in the context of models with minimal length; and in Ref.
\cite{vis}, in the framework of the Ho\v{r}ava gravity. Another motivation behind this work is the interest  in developing different probes into the properties of metamaterials. As an example,
spontaneous emission near hyperbolic metamaterials has been considered in Refs. \cite{kid,emit}.

The hyperbolic dispersion relation associated with nanostructured metamaterials and its application in engineering photonic density of states has been discussed in Ref. \cite{jac}. Here, we 
study the statistical mechanics of a photon gas with such dispersion relations for different wave-vector space topologies and obtain several new results. In particular, we show that they exhibit 
interesting properties such as the vanishing of density of states at certain frequencies and behaving like lower dimensional usual photon gases but with enhanced density of states.

In the next sections, after a brief review of the dispersion relation for the propagation of electromagnetic fields in an anisotropic medium, we show how hyperbolic dispersion relations could
arise as a result of negative permittivities. We then obtain an expression for the photonic density of states in a rather simple model.
We apply this to study the black body radiation and show that this results in novel properties such as exhibiting the behaviour of a photon gas in lower dimensional ordinary media. We study the
thermodynamics of such media by calculating the partition function in the grand canonical ensemble. We conclude by discussing the results.
\section{Hyperbolic dispersion relations}
The propagation of electromagnetic waves is governed by the well-known Maxwell equations
\begin{eqnarray}
{\boldmath\nabla}.{\bf D}&=&-\rho,\label{4a}\\
{\boldmath\nabla}\times{\bf E}&=&-\frac{\partial{\bf B}}{\partial t},\label{4b}\\
{\boldmath\nabla}.{\bf B}&=&0,\label{4c}\\
{\boldmath\nabla}\times{\bf H}&=&{\bf J}+\frac{\partial{\bf D}}{\partial t},\label{4d}
\end{eqnarray}
in which ${\bf D}={\boldmath\epsilon}\cdot{\bf E}$ and ${\bf B}={\boldmath\mu}\cdot{\bf H}$. In general, neglecting medium losses, the permittivity ${\boldmath\epsilon}$ and the
permeability ${\boldmath\mu}$ are symmetric tensors. Here, we consider an anisotropic dielectric medium with
\begin{equation}\label{l1}
{\boldmath\epsilon}=\left(\begin{array}{ccc}
\epsilon_x & 0 & 0 \\
0 & \epsilon_y & 0\\
0 & 0 & \epsilon_z
\end{array}\right),
{\boldmath\mu}=\left(\begin{array}{ccc}
\mu_0 & 0 & 0 \\
0 & \mu_0 & 0\\
0 & 0 & \mu_0
\end{array}\right)
\end{equation}
where $\epsilon_{x,y,z}$ are all constant, and seek plane wave solutions of the above equations in the absence of the electric charge $\rho$ and the current ${\bf J}$. Thus, we set
\begin{equation}\label{e1}
{\bf E}({\bf r},t)={\bf E}_0e^{i{\bf k}\cdot{\bf r}-i\omega(k)t}
\end{equation}
where the real part is to be taken. Inserting this back into the Maxwell equations (\ref{4b}) and (\ref{4d}) and making use of Eq. (\ref{4a}), one can show that the following relation holds
\begin{equation}\label{1c}
\frac{k^2_x}{k^2-\frac{\omega^2}{c^2}\varepsilon_x}+\frac{k^2_y}{k^2-\frac{\omega^2}{c^2}\varepsilon_y}+\frac{k^2_z}{k^2-\frac{\omega^2}{c^2}\varepsilon_z}=1
\end{equation}
in which $\varepsilon_i=\displaystyle{\frac{\epsilon_i}{\epsilon_0}}$ ($i=x,y,z$), and $c=\frac{1}{\sqrt{\epsilon_0\mu_0}}$ is the speed of light in vacuum. If we confine ourselves to the 
special case of a uniaxial medium with $\varepsilon_x=\varepsilon_y$, we can obtain the following simpler relation
\begin{equation}\label{1j}
\frac{k^2_x+k^2_y}{\varepsilon_z}+\frac{k^2_z}{\varepsilon_x}=\frac{\omega^2}{c^2}.
\end{equation}
For ordinary media, $\varepsilon_x$ and $ \varepsilon_z$ are positive and can be constant. However, for metamaterials, either of these can have negative values. Also, in general,
metamaterials are always highly dispersive media, i.e., they exhibit both temporal (frequency dependence) and spatial (wave-vector dependence) dispersions \cite{topm}. Thus, the explicit
forms of the permittivity components depend on the structure of the medium under consideration. Since the spatial dispersion effects are usually much weaker than the temporal ones, here, for 
convenience, we consider a dissipationless medium and ignore the spatial dispersion. In the presence of dissipation, which is the case for the metal-dielectric metamaterials, the permittivity 
components are complex and the above simple hyperbolic dispersion get replaced by hyperbolic-like dispersion relations \cite{lossy}.
However, we expect the results obtained here to be qualitatively valid for low-loss media as well.   

We consider a medium with
\begin{eqnarray}
\varepsilon_z&=&1\label{em1}\\
\varepsilon_x&=&1-\frac{\Omega^2}{\omega^2}\label{em2}
\end{eqnarray}
 in which $\Omega$ is some cutoff frequency. With this choice, we obtain negative values for $\varepsilon_x$ for frequencies under the cutoff value.
This characterizes a nonmagnetic metamaterial with negative $xx$ and $yy$ components of the permittivity. Thus, Eq. (\ref{1j}) reduces to
\begin{equation}\label{1h}
k^2_x+k^2_y-\frac{\omega^2}{\Omega^2-\omega^2}k^2_z=\frac{\omega^2}{c^2}.
\end{equation}
An example of such dispersion relations has been used in Ref. \cite{agran} (see also Ref. \cite{down}) to study the physics of vacuum in a strong magnetic field in which the permittivity 
components are $$\varepsilon_{xx}=\varepsilon_{yy}=\frac{1+\alpha}{1-\alpha}$$ and $$\varepsilon_{zz}=\alpha\left(1-\frac{\omega^2_s}{\omega^2}\right)$$
where $\omega_s$ is a constant frequency proportional to the inverse London penetration depth and $0\leq\alpha\leq 1$ is also a constant.

In this work, we confine ourselves to the case of nonmagnetic metamaterials. To take metamaterials with a magnetic response into account, one would be faced with a 
more complicated picture, in which a variety of mode-dependent dispersion relations (classified in Ref. \cite{depin}) should be considered. 
\section{The density of states}
\begin{figure}[h]
\centering
\includegraphics[scale=.55]{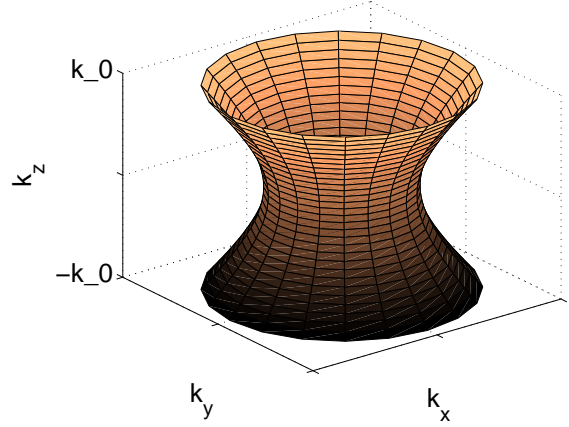}
\caption{The surface spanned by the dispersion relation $k^2_x+k^2_y-\frac{\omega^2}{\Omega^2-\omega^2}k^2_z=\frac{\omega^2}{c^2}$ in the wave-vector space. The density of states 
is given (up to a proportionality constant) by the volume enclosed by the hyperboloid and the planes $k_z=\pm k_0$. \label{fig1}}
\end{figure}

To study the statistical mechanics of the system, we first need to calculate the density of states. This can be obtained, as usual, from the wave-vector space volume
of the shell enclosed  by surfaces defined by $\omega$ and $\omega+d\omega$. Obviously, this volume is infinite, giving rise to a divergent density of states. However, in practice, there is a 
bound on the values of $k$ which is imposed by the physical properties of the medium. An example of this is discussed in Ref. \cite{jac} in which the metamaterial under study is fabricated via a 
special nanopatterning with a characteristic patterning scale $a$. This puts an upper cutoff value on $k$ which is of the order of $\frac{2\pi}{a}$. Thus, we need to calculate the above volume 
taking the constraint $k<k_c$ into account, $k_c$ being the cutoff value. The volume inside the hyperbolic surface described by Eq. (\ref{1h}) (depicted in Fig. \ref{fig1}) and the
planes $k_z=\pm k_0$, where $k_0\equiv\frac{\Omega^2-\omega^2}{c\Omega}$, equals
\begin{eqnarray*}
{\mathcal V}=\frac{2\pi}{3c^3\Omega^3}\omega^2(\Omega^2-\omega^2)(4\Omega^2-\omega^2)
\end{eqnarray*}
where we have used $k_cc=\Omega$. From this, we obtain
\begin{eqnarray*}
\frac{d{\mathcal V}}{d\omega}=\frac{4\pi}{3c^3\Omega^3}\omega(4\Omega^4-10\Omega^2\omega^2+3\omega^4).
\end{eqnarray*}

\begin{figure}[h]
\centering
\includegraphics[width=7cm]{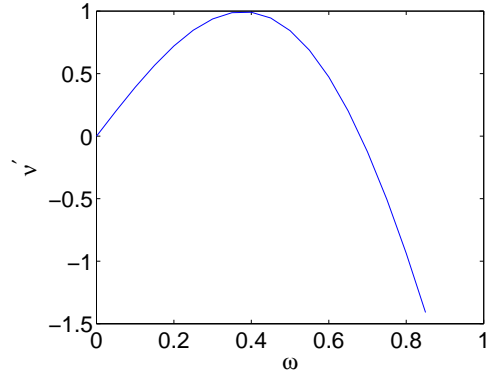}
\caption{The function $\nu^\prime=\frac{d\mathcal V}{d\omega}$ (in units of $\frac{4\pi\Omega^2}{3c^3}$) in terms of $\omega$ (in units of $\Omega$).
\label{fig2}}
\end{figure}

Basically, this gives the density of states in the interval $[\omega, \omega+d\omega]$. However, there are some exotic properties associated with this relation, namely, the appearance of 
negative values. This is because of the presence of a fixed cutoff value for $k$. In certain regions, by increasing $\omega$, the value of $k_0$ decreases so the net effect is a decrease in the
enclosed volume. The behaviour of $\frac{d\mathcal V}{d\omega}$ is depicted in Fig. \ref{fig2}.

The number of states in the interval between $\omega$ and $\omega+d\omega$ may be obtained by multiplying a factor of $\frac{2V}{8\pi^3}$ by the absolute value of
$\frac{d\mathcal V}{d\omega}d\omega$. Here, $V$ is the volume of the medium and the factor $\frac{2}{8}$ is inserted to account for two polarization states of each photon, and to include
the positive octant only. Thus, the density of states $\rho(\omega)$ is given by
\begin{equation}\label{1b}
\rho({\omega})d\omega=\frac{V}{3\pi^2c^3\Omega^3}\omega|4\Omega^4-10\Omega^2\omega^2+3\omega^4|d\omega.
\end{equation}
For $\omega>\Omega$, the medium obeys the usual spherical dispersion relation, and we have $\rho(\omega)d\omega=\frac{V}{\pi^2c^3}\omega^2d\omega$.

\begin{figure}[h]
\centering
\includegraphics[width=7cm]{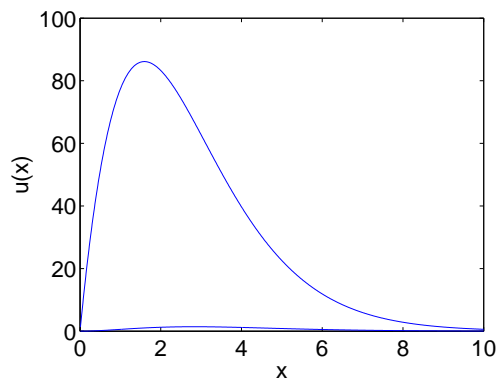}
\caption{Plot of the spectral density (in units of $\frac{1}{\pi^2\hbar^3\beta^4c^3}$) for $X=100$ (upper) compared to the result of the usual spherical dispersion relation (lower).
\label{fig2f}}
\end{figure}

An immediate effect of the above modified relation for the density of states can be seen by studying the black body radiation. To see this, we consider a metamaterial medium of the type 
described above and a gas of photons in equilibrium with the medium at a temperature $T=\frac{1}{k_B\beta}$ ($k_B$ being the Boltzmann constant). The energy density of the radiation 
emitted inside the medium at this temperature is given by the following expression
\begin{equation}\label{et1}
u(\omega)d\omega=\frac{1}{V}\frac{\hbar\omega\rho(\omega)d\omega}{e^{\beta\hbar\omega}-1}
\end{equation}
where $u(\omega)$ is the spectral density. For the usual spherical dispersion relation, this leads to the well-known Stefan-Boltzmann law. However, using the hyperbolic density of states, Eq.
(\ref{1b}), the result would be different. In practice, we need not to use the whole expression Eq. (\ref{1b}) here, but rather a simplified version would be sufficient. To see this, we first rewrite
the above expression in the following form
\begin{equation}\label{et1b}
u(x)=\frac{1}{3\pi^2c^3\hbar^3\beta^4}\frac{|4X^4-10X^2x^2+3x^4|}{X^3}\frac{x^2}{e^x-1}
\end{equation}
where $x=\beta\hbar\omega$ and $X=\beta\hbar\Omega$. Now, we note that for a medium with nanoscale characteristic patterning length
$a\sim 10^{-8} m$, for room temperatures we have $X\sim 10^{3}$, which means that we can practically confine ourselves to the region $x\ll X$. Thus, we obtain
\begin{equation}\label{et1c}
u(x)=\frac{4X}{3\pi^2c^3\hbar^3\beta^4}\frac{x^2}{e^x-1}
\end{equation}
The functional form of this expression resembles its counterpart for a usual two-dimensional medium. This might be explained from the dispersion relation Eq. (\ref{1h}) where in the left-hand
side we have something like the norm of a vector computed with respect to a metric with two spacelike components  and a timelike one. Also, compared with a two dimensional photon gas, here
the spectral density contains an extra $\frac{4X}{3c\hbar\beta}$ factor, which is of the order of inverse characteristic length $a^{-1}$, leading to a huge enhancement of the density. The
above relation is plotted in Fig. \ref{fig2f} which shows radical enhancement of the spectral density compared with a usual photon gas in $3+1$ Minkowskian space.

\begin{figure}[h]
\centering
\includegraphics[width=7cm]{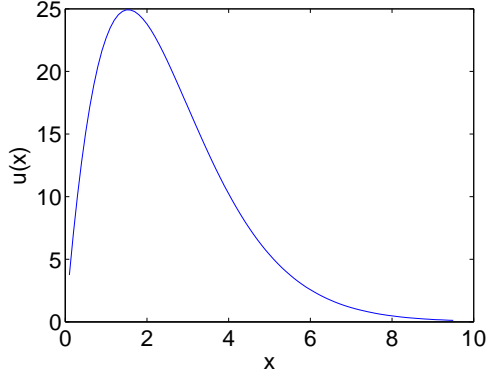}
\caption{Spectral density (in units of $\frac{1}{\pi^2\hbar^3\beta^4c^3}$) for $X=10$.\label{fig2g}}
\end{figure}

In spite of the above arguments, one may still be interested in studying the behaviour of the spectral density for situations where, say,  $X\sim 10^1$, corresponding to a microscale patterning
length $a\sim 10^{-6}m$. This is depicted in Fig. \ref{fig2g}. Qualitatively, this looks like the previous case.
\section{The partition function}
The thermodynamics of the above described photon gas can be derived from the partition function in the appropriate limit. Considering a grand canonical ensemble for
the photon gas, we can write the grand partition function as follows
\begin{equation}\label{eq1}
\ln{\mathcal Q}\equiv\frac{PV}{k_BT}=-\int_{0}^{\infty}\rho(\omega)\ln(1-e^{-\beta\hbar\omega})d\omega
\end{equation}
Inserting Eq. (\ref{1b}) into the above expression, we get
\begin{eqnarray}
\ln{\mathcal Q}&=&\left\{
\frac{-V}{3\pi^2c^3\Omega^3}\left(\int_0^{\alpha\Omega}-\int_{\alpha\Omega}^\Omega\right)\omega(4\Omega^4-10\Omega^2\omega^2\right.\nonumber\\&&\left.+3\omega^4)
-\frac{V}{\pi^2c^3}\int_\Omega^\infty\omega^2\right\}\ln(1-e^{-\beta\hbar\omega})d\omega
\end{eqnarray}
where $\alpha\approx 0.67$ is the root of $4-10x^2+3x^4=0$. Now, using our previous argument on largeness of $X=\beta\hbar\Omega$ for a nanopatterned metamaterial, we can ignore the 
last two integrals in the above expression. This leaves us with
\begin{eqnarray}
\ln{\mathcal Q}&=&\frac{-V\Omega}{3\pi^2c^3\beta^2\hbar^2}\int_0^{\alpha X}f(x)dx
\end{eqnarray}
in which $$f(x)=x\left(4-10\frac{x^2}{X^2}+3\frac{x^4}{X^4}\right)\ln(1-e^{-x}).$$
In view of the largeness of $X$, this can be approximated as
\begin{eqnarray}
\ln{\mathcal Q}&\approx&\frac{V\Omega}{6\pi^2c^3\beta^2\hbar^2}\int_0^\infty \frac{4x^2-\frac{5x^4}{X^2}+\frac{x^6}{X^4}}{e^x-1}dx.
\end{eqnarray}
Inserting the values of the Bose-Einstein integrals above, this reduces to
\begin{eqnarray}
\ln{\mathcal Q}&\approx&\frac{4V\Omega}{3\pi^2c^3\beta^2\hbar^2}\left(\zeta(3)-\frac{15}{X^2}\zeta(5)+\frac{90}{X^4}\zeta(7)\right)
\end{eqnarray}
where $\zeta$ is the Riemann zeta function.

The internal energy is given by
\begin{eqnarray}
U&=&-\frac{\partial}{\partial\beta}\ln{\mathcal Q}\nonumber\\&\approx&
\frac{8V\Omega}{3\pi^2c^3\beta^3\hbar^2}\left(\zeta(3)-\frac{30}{X^2}\zeta(5)+\frac{270}{X^4}\zeta(7)\right)
\end{eqnarray}
which resembles the energy of a usual photon gas in two dimensions together with correction terms of the form $X^{-2}$ and $X^{-4}$ which are very small for a metamaterial medium with
nanoscale patterning length. Comparing the last two equations, one reaches the following relation
\begin{equation}\label{e20}
P=\frac{1}{2}\frac{U}{V}\left(1-O(X^{-2})\right)
\end{equation}
in which higher order correction terms are neglected. Other thermodynamical quantities such as entropy can also be obtained from the above partition function. The result of these calculations
is consistent with the result obtained in the previous section according to which the photon gas under consideration behaves like an ordinary photon gas in $2+1$ dimensions with some small 
corrections.

\begin{figure}[h]
\centering
\includegraphics[scale=.6]{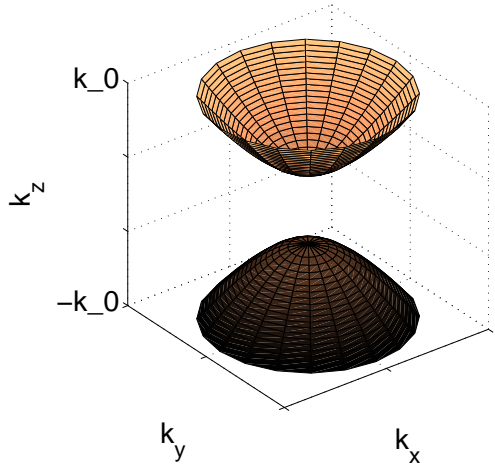}
\caption{The surface spanned by the dispersion relation $k^2_x+k^2_y-\left(\frac{\Omega^2}{\omega^2}-1\right)k^2_z=-\frac{\Omega^2-\omega^2}{c^2}$.\label{fig77}}
\end{figure}

It would also be interesting to consider a metamaterial with $\varepsilon_x=\varepsilon_y>0$ and $\varepsilon_z <0$. This can be modelled by interchanging $\varepsilon_x$ and 
$\varepsilon_z$ in Eqs. (\ref{em1}) and (\ref{em2}). The relevant dispersion relation is then
\begin{equation}\label{em3}
k^2_x+k^2_y-\left(\frac{\Omega^2}{\omega^2}-1\right)k^2_z=-\frac{\Omega^2-\omega^2}{c^2}.
\end{equation}
In this case, the wave-vector space configuration is again a hyperboloid, but with a different topology, namely a two sheeted hyperboloid depicted in Fig. \ref{fig77}.

\begin{figure}[h]
\centering
\includegraphics[width=7cm]{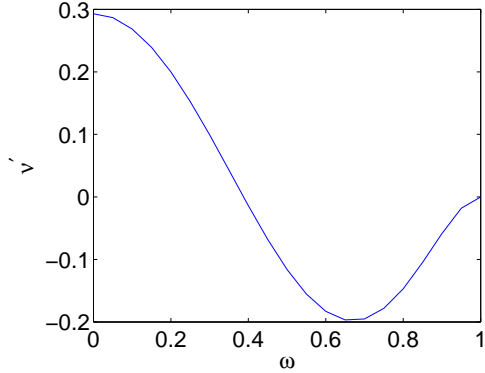}
\caption{The function $\nu^\prime=\frac{d\mathcal V}{d\omega}$ (in units of $\frac{4\pi\Omega^2}{3c^3}$) in terms of $\omega$ (in units of $\Omega$) for the surface
$k^2_x+k^2_y-\left(\frac{\Omega^2}{\omega^2}-1\right)k^2_z=-\frac{\Omega^2-\omega^2}{c^2}$.\label{fig66}}
\end{figure}

In this case, we are interested in the volume between the surfaces spanned by the relation Eq. (\ref{em3}) and the cutoff planes $k_z=\pm k_0=\frac{\omega}{\Omega c}
{\sqrt{2\Omega^2-\omega^2}}$. Again, this volume can decrease or increase with increasing values of $\omega$. The behaviour of the derivative of this volume with respect to frequency is
shown in Fig. \ref{fig66}.

Repeating the same steps as in the previous sections, for $\omega<\Omega$ we reach
\begin{eqnarray}\label{e77}
\rho(\omega)&=&\frac{V}{3\pi^2c^3\Omega^3\sqrt{2\Omega^2-\omega^2}}\left|-\Omega^6+\Omega^4\omega^2+5\Omega^2\omega^4\right.\nonumber\\&&
\left.-3\omega^6+(\Omega^5-3\Omega^3\omega^2){\sqrt{2\Omega^2-\omega^2}}\right|
\end{eqnarray}
This can be approximated as
\begin{equation}\label{e77a}
\rho(\omega)d\omega\approx \frac{V}{3\pi^2c^3}\left(\frac{\sqrt{2}-1}{\sqrt{2}}\Omega^2+\frac{\sqrt{2}-6}{2}\omega^2\right)d\omega
\end{equation}
This shows a constant density of states corrected with a $\omega^2$ term. One may interpret this as the density of states of a photon gas in $1+1$ dimensions for
large values of $\Omega$.
\section{Discussion and conclusions}\label{phys}
We have considered a photonic gas with modified dispersion relation propagating in a nonmagnetic metamaterial medium with some negative components of the permittivity tensor. From the 
hyperbolic dispersion relations, we obtained the density of states, which in the case where the $xx$ and $yy$ components of the permittivity tensor are negative, shows some
interesting features such as vanishing at a certain frequency. For the case where only the $zz$ component is negative, the surface  spanned by the dispersion relation in the wave-vector space is
still hyperbolic but with a different topology and the density of states can decrease or increase with frequency. We studied the thermodynamics of the system and showed that it resembles a
photonic gas in an ordinary one or two dimensional medium but with enhanced density of states.  The results are of interest in studying the behaviour of physical systems in spacetimes with non-
Lorentzian metric signatures. The relevant calculations are performed under the assumption that the medium dielectric coefficients have a plasma-like frequency dependence but are independent 
of the wave-vector. It would be interesting to consider more general situations where these are functions of both the frequency and the wave-vector.
\section*{Acknowledgements}
I would like to thank the Abdus Salam ICTP where part of this work was done. I also thank two anonymous referees of Journal of Optics for comments.

\end{document}